\RequirePackage{lineno} 
\documentclass[aps,prc,twocolumn,showpacs,superscriptaddress,groupedaddress, ]{revtex4}  
\usepackage{graphicx}  
\usepackage{dcolumn}  
\usepackage{bm}           
\usepackage{amssymb}  
\usepackage{multirow}  

\begin{document}
\widetext
\leftline{ \today}
\title{Azimuthal Emission Patterns of $K^{+}$ and of $ K^{-} $ Mesons in Ni + Ni Collisions 
near the Strangeness Production Threshold}

\def\heid{Physikalisches Institut der Universit\"{a}t Heidelberg, Heidelberg, Germany}
\def\darm{GSI Helmholtzzentrum f\"{u}r Schwerionenforschung GmbH, Darmstadt, Germany}
\def\seou{Korea University, Seoul, Korea}
\def\cler{Laboratoire de Physique Corpusculaire, IN2P3/CNRS, and Universit\'{e} Blaise Pascal, Clermont-Ferrand, France}
\def\zagr{Ru{d\llap{\raise 1.22ex\hbox{\vrule height 0.09ex width 0.2em}}\rlap{\raise 1.22ex\hbox{\vrule height 0.09ex width 0.06em}}}er Bo\v{s}kovi\'{c} Institute, Zagreb, Croatia}
\def\munI{Excellence Cluster Universe, Technische Universit\"{a}t M\"{u}nchen, Garching, Germany}
\def\munII{E12, Physik Department, Technische Universit\"{a}t M\"{u}nchen, Garching, Germany}
\def\vien{Stefan-Meyer-Institut f\"{u}r subatomare Physik, \"{O}sterreichische Akademie der Wissenschaften, Wien, Austria}
\def\sp{University of Split, Split, Croatia}
\def\buda{Wigner RCP, RMKI, Budapest, Hungary}
\def\wars{Institute of Experimental Physics, Faculty of Physics, University of Warsaw, Warsaw, Poland}
\def\mosc{Institute for Theoretical and Experimental Physics, Moscow, Russia}
\def\dres{Institut f\"{u}r Strahlenphysik, Helmholtz-Zentrum Dresden-Rossendorf, Dresden, Germany} 
\def\harb{Harbin Institute of Technology, Harbin, China}
\def\kurc{Kurchatov Institute, Moscow, Russia}
\def\buch{Institute for Nuclear Physics and Engineering, Bucharest, Romania}
\def\stra{Institut Pluridisciplinaire Hubert Curien and Universit\'{e} de Strasbourg, Strasbourg, France}
\def\tsing{Department of Physics, Tsinghua University, Beijing 100084, China}
\def\lan{Institute of Modern Physics, Chinese Academy of Sciences, Lanzhou, China}

\author{V.~Zinyuk}\email{v.zinyuk@gsi.de} \affiliation{\heid}
\author{T.I.~Kang}\email{t.i.kang@gsi.de}  \affiliation{\darm}\affiliation{\seou}
\author{Y.~Leifels} \affiliation{\darm}
\author{N.~Herrmann} \affiliation{\heid}
\author{B.~Hong} \affiliation{\seou}
\author{R.~Averbeck} \affiliation{\darm}
\author{A.~Andronic} \affiliation{\darm} 
\author{V.~Barret} \affiliation{\cler} 
\author{Z.~Basrak} \affiliation{\zagr} 
\author{N.~Bastid} \affiliation{\cler}
\author{M.L.~Benabderrahmane} \affiliation{\heid}
\author{M.~Berger} \affiliation{\munI} \affiliation{\munII}
\author{P.~Buehler} \affiliation{\vien} 
\author{M.~Cargnelli} \affiliation{\vien} 
\author{R.~\v{C}aplar} \affiliation{\zagr}
\author{I.~Carevic} \affiliation{\sp} 
\author{P.~Crochet} \affiliation{\cler} 
\author{I.~Deppner} \affiliation{\heid}
\author{P.~Dupieux} \affiliation{\cler}
\author{M.~D\v{z}elalija} \affiliation{\sp}
\author{L.~Fabbietti} \affiliation{\munI} \affiliation{\munII}
\author{Z.~Fodor} \affiliation{\buda}
\author{P.~Gasik} \affiliation{\wars}
\author{I.~Ga\v{s}pari\'c} \affiliation{\zagr}
\author{Y.~Grishkin} \affiliation{\mosc}
\author{O.N.~Hartmann} \affiliation{\darm}
\author{K.D.~Hildenbrand} \affiliation{\darm}
\author{J.~Kecskemeti} \affiliation{\buda}
\author{Y.J.~Kim} \affiliation{\darm}
\author{M.~Kirejczyk} \affiliation{\wars}
\author{M.~Ki\v{s}} \affiliation{\darm} \affiliation{\zagr}
\author{P.~Koczon} \affiliation{\darm}
\author{R.~Kotte} \affiliation{\dres}
\author{A.~Lebedev} \affiliation{\mosc}
\author{A.~Le F\`{e}vre} \affiliation{\darm}
\author{J.L.~Liu} \affiliation{\heid} \affiliation{\harb}
\author{X.~Lopez} \affiliation{\cler}
\author{V.~Manko} \affiliation{\kurc}
\author{J.~Marton} \affiliation{\vien}
\author{T.~Matulewicz} \affiliation{\wars}
\author{R.~M\"{u}nzer} \affiliation{\munI} \affiliation{\munII}
\author{M.~Petrovici} \affiliation{\buch}
\author{K.~Piasecki} \affiliation{\wars}
\author{F.~Rami} \affiliation{\stra}
\author{A.~Reischl} \affiliation{\heid}
\author{W.~Reisdorf} \affiliation{\darm}
\author{M.S.~Ryu} \affiliation{\seou}
\author{P.~Schmidt} \affiliation{\vien}
\author{A.~Sch\"{u}ttauf} \affiliation{\darm}
\author{Z.~Seres} \affiliation{\buda}
\author{B.~Sikora} \affiliation{\wars}
\author{K.S.~Sim} \affiliation{\seou}
\author{V.~Simion} \affiliation{\buch}
\author{K.~Siwek-Wilczy\'{n}ska} \affiliation{\wars}
\author{V.~Smolyankin} \affiliation{\mosc}
\author{K.~Suzuki} \affiliation{\vien}
\author{Z.~Tyminski} \affiliation{\wars}
\author{P.~Wagner} \affiliation{\stra}
\author{E.~Widmann} \affiliation{\vien}
\author{K.~Wi\'{s}niewski} \affiliation{\heid} \affiliation{\wars}
\author{Z.G.~Xiao} \affiliation{\tsing}
\author{I.~Yushmanov} \affiliation{\kurc}
\author{Y.~Zhang} \affiliation{\heid} \affiliation{\lan}
\author{A.~Zhilin} \affiliation{\mosc}
\author{J.~Zmeskal} \affiliation{\vien}

\collaboration{FOPI Collaboration} \noaffiliation
\author{E.~Bratkovskaya} 
\affiliation{Frankfurt Institute for Advanced Studies, Frankfurt am Main, Germany}
\affiliation{Institute for Theoretical Physics, Johann Wolfgang Goethe Universität,  Frankfurt am Main, Germany}
\author{C.~Hartnack}  
\affiliation{SUBATECH, UMR 6457, Ecole des Mines de Nantes - IN2P3/CNRS - Universit\'e de Nantes, Nantes, France}

\begin{abstract}
Azimuthal emission patterns of $K^\pm$ mesons have been measured in
Ni + Ni collisions with the FOPI spectrometer at a beam kinetic energy of 1.91\,A GeV. 
The transverse momentum $p_{T}$ integrated directed and elliptic flow  of
$K^{+}$ and $K^{-}$ mesons 
as well as the centrality dependence of $p_{T}$\,-\,differential  directed flow
of $K^{+}$ mesons are compared to the predictions of HSD and IQMD transport models.
The data exhibits different propagation patterns 
of $K^{+}$ and $K^{-}$ mesons in the compressed and heated
nuclear medium and favor the existence of a kaon-nucleon in-medium potential, repulsive 
for $K^{+}$ mesons and attractive for $K^{-}$ mesons. 
\end{abstract}
\maketitle
\section{Introduction}
Relativistic heavy-ion collisions at bombarding energies of
1\,-\,2\,A GeV provide the unique possibility to study nuclear
matter at high temperatures, around 100~MeV,  and baryon densities
about 2\,-\,3 times the normal nuclear matter density($ \rho_{0} $) \cite{Fuchs.2006,B.Hong.prc.1998}.
Under these conditions, the properties of hadrons may be altered as a result of various non-trivial 
in-medium effects like the partial restoration of the spontaneously broken chiral symmetry, 
the modified baryon-meson couplings, and the nuclear potential.
Whether and how hadronic properties, such as masses, widths and dispersion
relations are modified in the hot and dense nuclear medium 
is a topic of great current interest.
In particular, strange mesons produced around the production threshold energies in 
nucleon\,-\,nucleon collisions are considered 
to be sensitive to in-medium modifications.
Various theoretical approaches agree qualitatively 
predicting  slightly repulsive  $KN$\,-  and strongly attractive
 $\overline{K}N$ potentials \cite{Kaplan.1986}.
The depth of the  $\overline{K}N$ potential at finite densities is,
however, not well constrained by currently available data and is a matter of 
an active theoretical dispute \cite{Hartnack2011}.  
If the  $K^{-}N$ potential is sufficiently deep, this might have 
exciting consequences for the stability of neutron stars  \cite{li.1997} or 
for the existence of  deeply bound $K^{-}$ states \cite{Akaishi.2002}.

Heavy-ion experiments, with the capability to identify kaons and antikaons,
have been performed with the KaoS, FOPI and HADES detector systems 
at the heavy-ion synchrotron (SIS) of GSI, aiming at measuring the in-medium properties. 
A significantly enhanced yield of $K^{-}$ mesons
relative to that of $K^{+}$ \cite{Best.1997, Laue.prl.1999}, 
an increase of the $K^{-}/K^{+}$ ratio at low kinetic energy of 
kaons \cite{Laue.prl.1999, K.Wis.2000},  and different freeze-out conditions of $K^{+}$ and $K^{-}$ mesons
were observed \cite{Foester.2007}, the latter at least partially explained by
the production of $\phi$ mesons \cite{Hades2009}. 
After the suggestion that the $KN$ potential should
manifest itself in the collective motion of kaons, referred to as `flow' \cite{G.Q.Li.prl.1995}, 
a lot of effort was invested to deduce the strength of the kaon potential 
by measuring the kaon flow in heavy-ion collisions
\cite{Ritman.zpa.1995, Shin.1998, Crochet.2000, Uhlig.2005}.
Experimental difficulties due to the small production rate of $ K^{-} $ mesons
in comparison to $ K^{+} $ in the SIS energy regime restrict the measurements. 
Currently available flow results for $K^{-}$ mesons are not sufficient to            
draw conclusions on the existence and strength of the  $\overline{K}N$ in-medium potential.

In this article, we report for the first time on the  
simultaneous measurements of $ K^{+}$ and $K^{-} $ mesons 
with a large acceptance spectrometer at an incident beam energy of 1.91\,AGeV 
that is close to the various strangeness production threshold energies.
The data are compared to state-of-the-art transport
calculations with and without  the assumption of an in-medium potential.
In particular, we show the azimuthal anisotropy of $K^{-}$ mesons
in a wide range of rapidity and the centrality dependence
of $ p_{T} $ differential flow of $ K^{+} $ mesons.

\section{Data accumulation and analysis}
The experiment was performed with the FOPI spectro\-meter,  an azimuthally symmetric apparatus 
comprising several sub detectors \cite{FOPI.det}. 
Recently a high resolution highly granular time-of-flight (ToF) Barrel based on 
Multi-strip Multi-gap Resistive Plate Counters (MMRPC) \cite{kis.2011}
was added to the FOPI apparatus improving significantly the kaon identification capability. 
Charged kaons are identified based on the  ToF-information from MMRPC
 and Plastic Scintillator Barrel (PSB), combined with the momentum
information from the Central Drift Chamber (CDC, $27^{\circ} < \theta_{lab} <113^{\circ}$ ), 
see Table~\ref{table:accep}.
 {\renewcommand{\arraystretch}{1.4}
 \renewcommand{\tabcolsep}{0.1cm}
\begin{table}[!h]
\begin{ruledtabular}
\caption{\label{table:accep} 
 Charged kaon acceptance in terms of laboratory 
  momentum and polar angle are shown with corresponding  
  signal-to-background-ratios (S/B) (see text for details).} 
\begin{tabular}{c c c c c c }
 ToF &$\theta_{lab}$ & \multicolumn{2}{c}{$K^+$} & \multicolumn{2}{c}{$K^-$} \\ \cline{3-6}
& (deg)  & $p_{lab}$ (GeV/$c$) & S/B & $p_{lab}$ (GeV/$c$) & S/B\\ \hline
 \multicolumn{1}{c}{MMRPC}& [30, 55] &  [0.13, 0.9] & $>22$ & [0.13, 0.7] & $>8$   \\ 
 \multicolumn{1}{c}{PSB}& [55, 110] &  [0.13, 0.55] & $>10$ & [0.13, 0.45] & $>4$  
\end{tabular} 
\end{ruledtabular}
\end{table}
}
 \begin{figure}[!t]
 \includegraphics[width=.95\columnwidth]{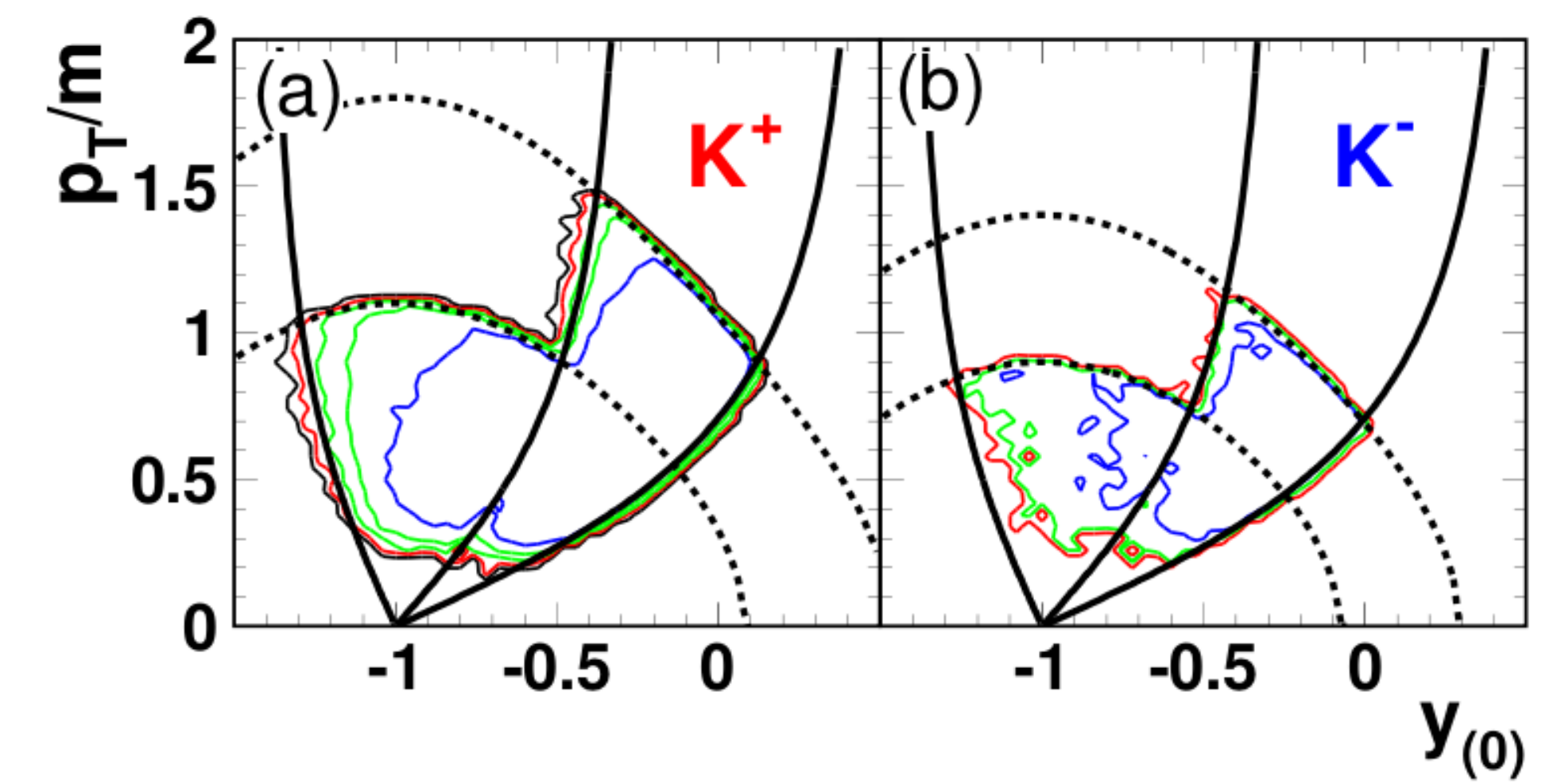}
 \includegraphics[width=.95\columnwidth]{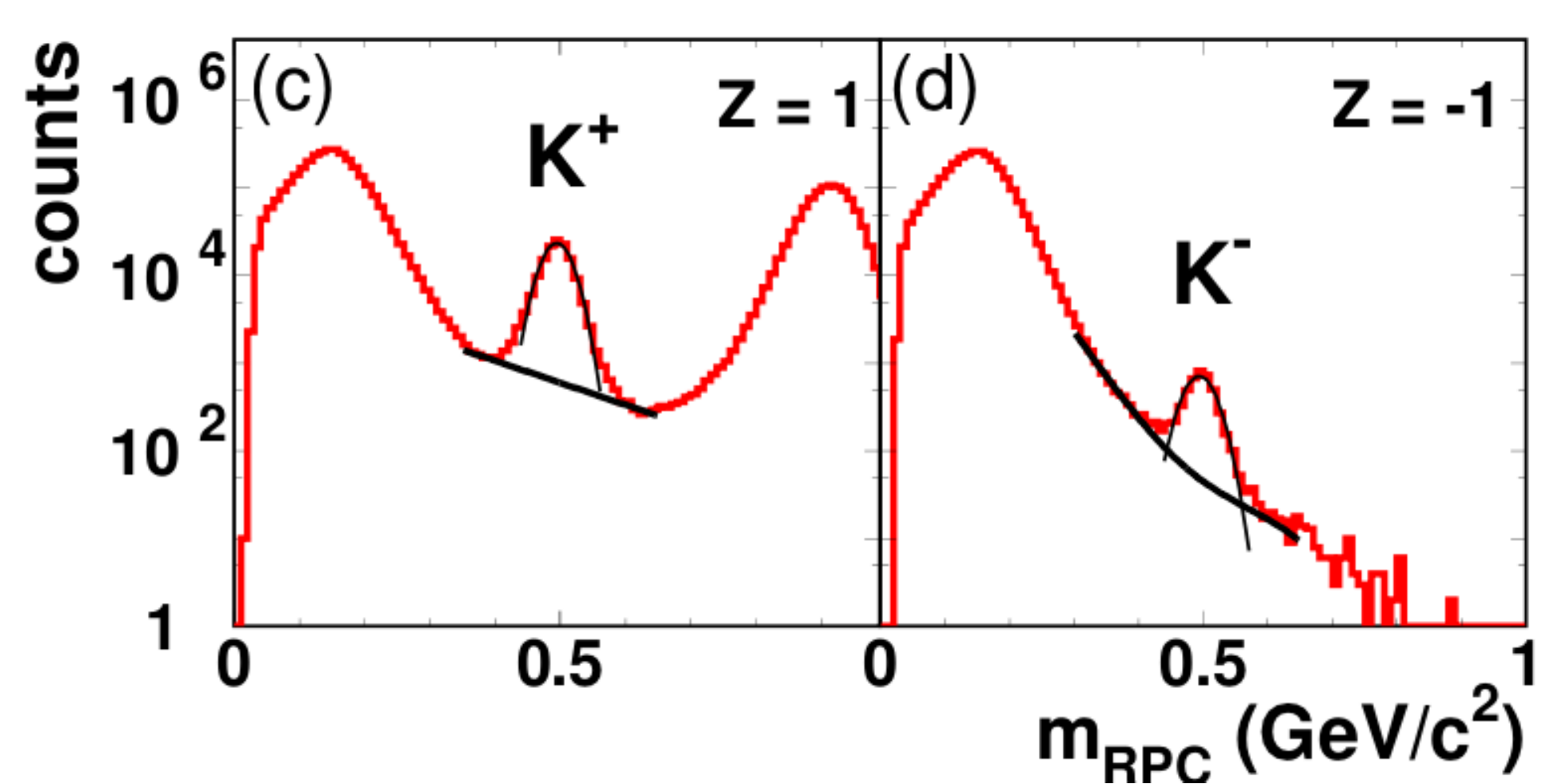}
 \caption{\label{fig:accep.bmk} (color online) 
 Upper panel: Measured yield of $K^{+}$  and $K^{-}$  mesons: 
  $ p_{T}/m_K $ as a function of $ y_{(0)} $. 
 The contour levels correspond to logarithmically increasing intensity.
 The solid curves denote the geometrical limits of the detector 
 acceptance ($ \theta_{lab} = 30^\circ$, $55^\circ$ and $ 110^\circ$). 
 The dashed curves corresponds to $p_{lab} = 0.55$ and 0.9 GeV/$c$
 for $ K^{+} $ (left) and $p_{lab} = 0.45$ and 0.7 GeV/$c$ for $ K^{-} $ (right).
 Lower panel: The mass spectra from MMRPC for Z~=~1 and -1. 
 The solid lines represent Gaussian fit functions for the signal and
 exponential functions for the background (see text for details).
  }
 \end{figure}

The acceptance range of the detector for $ K^{\pm} $
is shown in the upper panel of Fig.~\ref{fig:accep.bmk} 
in terms of normalized transverse momentum, $p_{T}/m$, and normalized rapidity, 
$y_{(0)}$ = $y_{lab}$/$y_{cm}-1$, defined to be +1 (-1) at projectile (target) rapidity.
About $69\times10^{6}$ events were recorded,
triggering on the most central 60\% of the total geometrical cross section
($\sigma_{trig}=1.6\,$ b). 
In total, 233,300 $K^+$ and 5,200 $K^-$ mesons 
were identified within  2$ \sigma $ 
around the fitted signal peaks from ToF mass spectra
(as visualized for the MMRPC in Fig.~\ref{fig:accep.bmk}, lower panel).
In order to account for the contamination  from 
pions and protons as well as misidentified tracks
the background distribution was estimated in a worst-case scenario, i.e.
as linear background connecting the minima around the fitted signal peaks in the mass spectra.
Following these defintions for `signal' and `background' 
the signal-to-background-ratios (S/B), as listed in Table~\ref{table:accep}, were estimated.
The events were sorted into different centrality intervals by imposing conditions on the baryon 
multiplicity (Mul), shown in Table~\ref{table:rpcor}. The baryon multiplicity
contains all charged particles from the Plastic Wall ($6.5^{\circ} < \theta_{lab} <23^{\circ}$) 
and p, d, t, $^3$He and $^4$He from the CDC.
The reaction plane was reconstructed event-wise by the 
transverse momentum method \cite{danielewicz} including all
particles except identified pions and kaons within the CDC and Plastic Wall acceptance
 outside the mid-rapidity interval $|y_{(0)}|<0.3$.
The particle of interest was excluded in order to avoid autocorrelations. 

The phenomenon of collective flow \cite{Herrmann.1999} can be quantitatively described 
in terms of anisotropies of the azimuthal emission pattern, 
expressed by a Fourier series:
\begin{equation}
\label{EQ:Fourier}
\frac{dN}{d\phi} \propto \left( 1+ 2v_{1}\cos(\phi)+2v_{2}\cos(2\phi)+ ...\right),
\end{equation}
where $\phi$ is the  azimuthal angle of the outgoing particle with respect to
the reaction plane \cite{Voloshin.1996}.
The first order Fourier coefficient, $v_{1}$,  
describes the collective sideward deflection of particles in the reaction plane, called `directed flow'. 
The second order Fourier coefficient, $v_{2}$,
describes the emission pattern in- versus out- of the reaction plane, referred to as `elliptic flow'
\cite{Andronic,Crochet.2000}.
The Fourier coefficients are corrected for the accuracy 
of the reaction plane determination according to the Ollitrault 
method \cite{Ollitrault.1998}. 
The mean correction values, $f_{1}$ and $f_{2}$ (given in Table~\ref{table:rpcor}) are calculated 
separately for each evaluated centrality(multiplicity) interval. 
The correction values $f_{n}$ are applied to the measured $v_{n}$ values according to the event's multiplicity.
Possible resolution effects due to wide centrality bins, as discussed in \cite{Schmah.2012}, were investigated and found 
to be negligible for the present analysis. 
Note that the most peripheral 
events (Mul $<$ 20) were rejected to assure a minimal accuracy of the reaction plane determination.
{\renewcommand{\arraystretch}{1.4}
\renewcommand{\tabcolsep}{0.1cm}
\begin{table}[h]
\begin{ruledtabular}
\caption{\label{table:rpcor}
Definition of event classes: 
(a) total, (p) peripheral  and (c) central events. 
The corresponding cross section  $\sigma$, mean impact 
parameter $\langle b \rangle$, the r.m.s. of  b distribution:  $ \Delta b$  and 
the reaction plane correction factors $f_{1}$ for $v_{1}$ and $f_{2}$ 
for  $v_{2}$ are listed.}
\begin{center}
\begin{tabular}{c c c c c c }
  &Mul & $\sigma$ (b) & $\langle b \rangle \pm \Delta b$ (fm)  & $f_{1}$  & $f_{2}$ \\  \hline
 (a) &$[20, 90]$ & $1.09 \pm 0.10$ &  $3.90 \pm 1.41$& $1.5\pm 0.1$ & $3.0 \pm 0.1$ \\  
 (p) & $[20, 48]$ & $0.79 \pm 0.05$&$4.54 \pm 0.95 $& $ 1.5 \pm 0.1$ & $3.0 \pm 0.1$ \\ 
 (c) &$[49, 90]$ & $0.30 \pm 0.05$& $2.11 \pm 0.80 $& $1.6  \pm 0.1 $&  $3.1 \pm 0.2$  \\
\end{tabular}
\end{center}
\end{ruledtabular}
\end{table}
}
\section{Results}
\begin{figure*}
\begin{center}
 \includegraphics[width=0.9\textwidth]{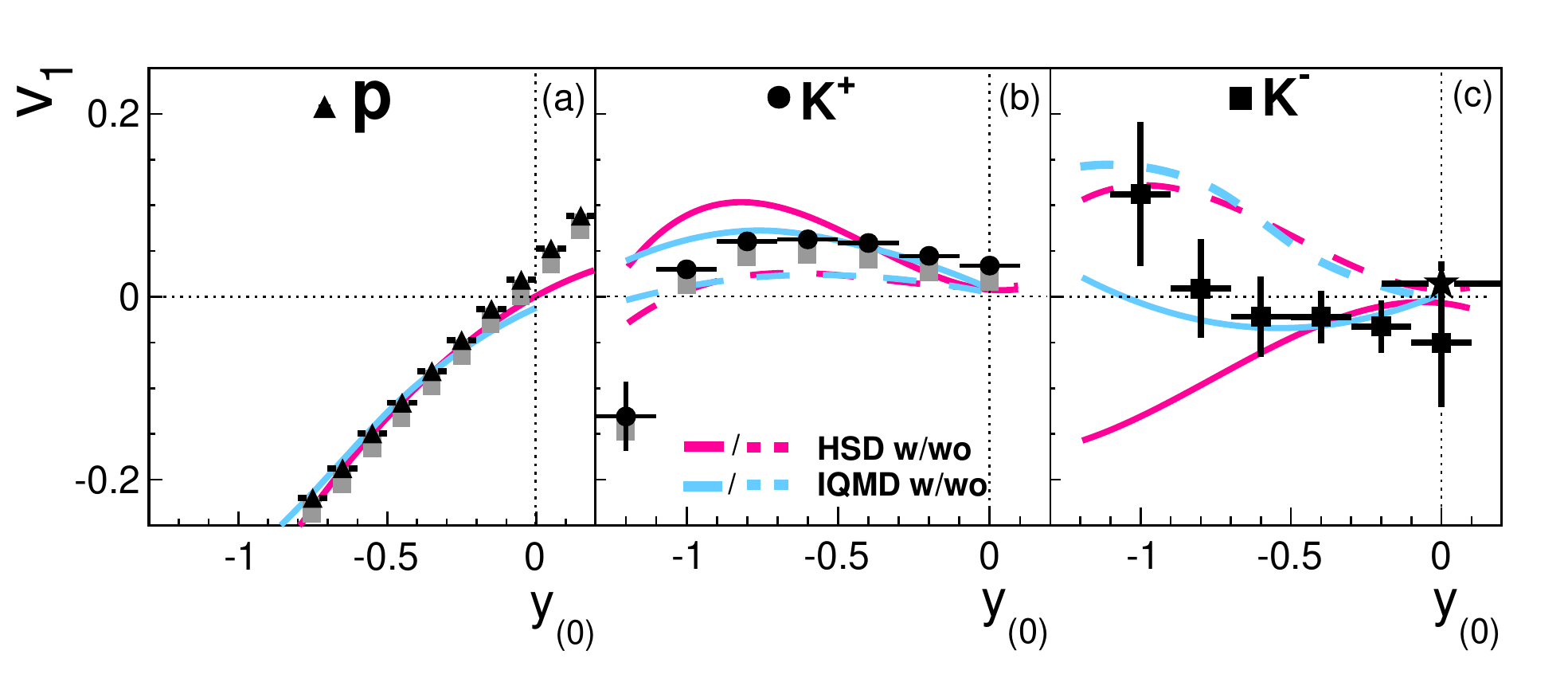}
\caption{\label{fig:v1} (color online)
Rapidity dependence of $v_{1}$ for protons (a), $ K^{+} $ (b) and $ K^{-} $ mesons (c), 
in comparison to HSD  and IQMD with (`w') and without (`wo') 
in-medium potential.
Error bars (boxes) denote statistical (systematic) uncertainties. 
The star symbols for $K^{-}$ mesons at mid-rapidity in (c)
are from the high statistics data in the range $p<1.0$ GeV/$c$ with S/B$>5$.}
\end{center}
\end{figure*}

\begin{figure*}
\begin{center}
\includegraphics[width=0.9\textwidth]{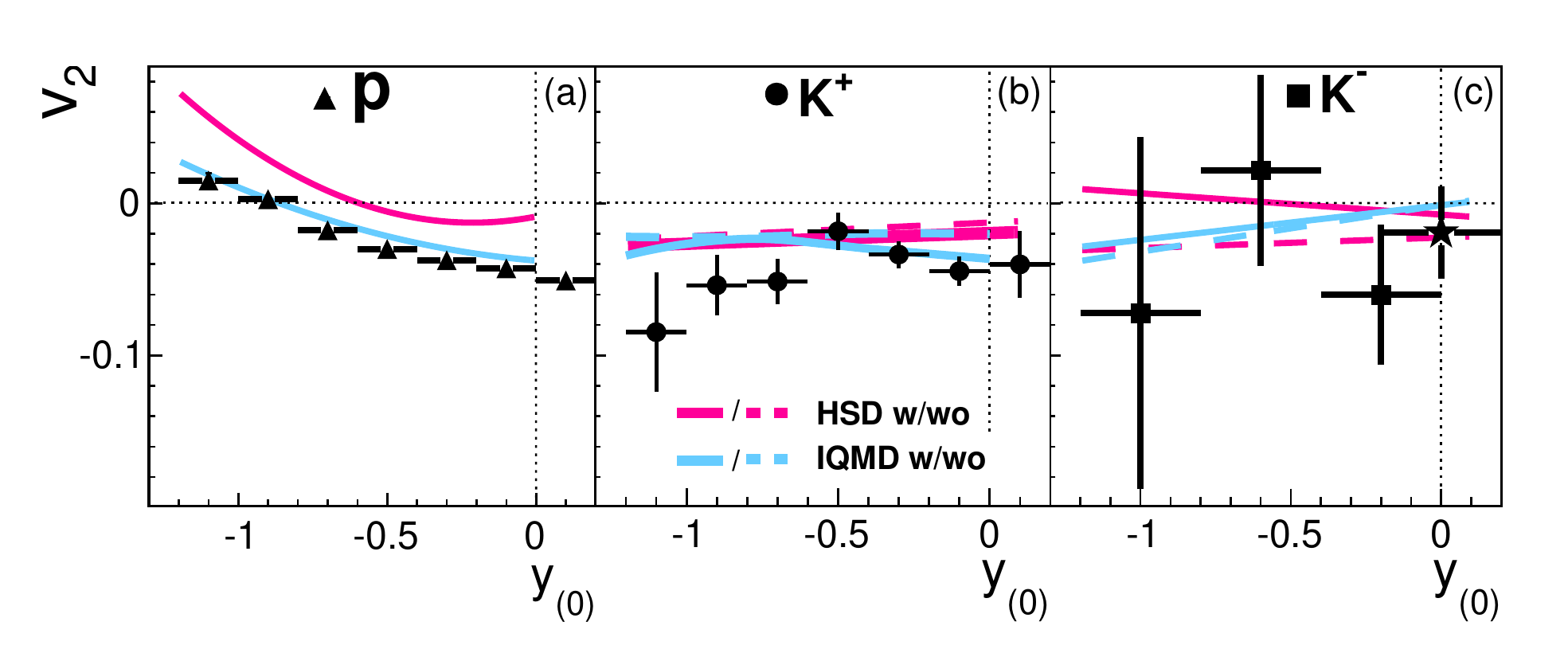}
\caption{\label{fig:v2} (color online)
Rapidity dependence of $v_{2}$ for  protons (a), $ K^{+} $ (b) and $ K^{-} $ mesons (c),
 in comparison to HSD  and IQMD transport model predictions. Lines and symbols as 
in Fig.~\ref{fig:v1}.}
\end{center}
\end{figure*}

\begin{figure}[t]
 \includegraphics[width=.99\columnwidth]{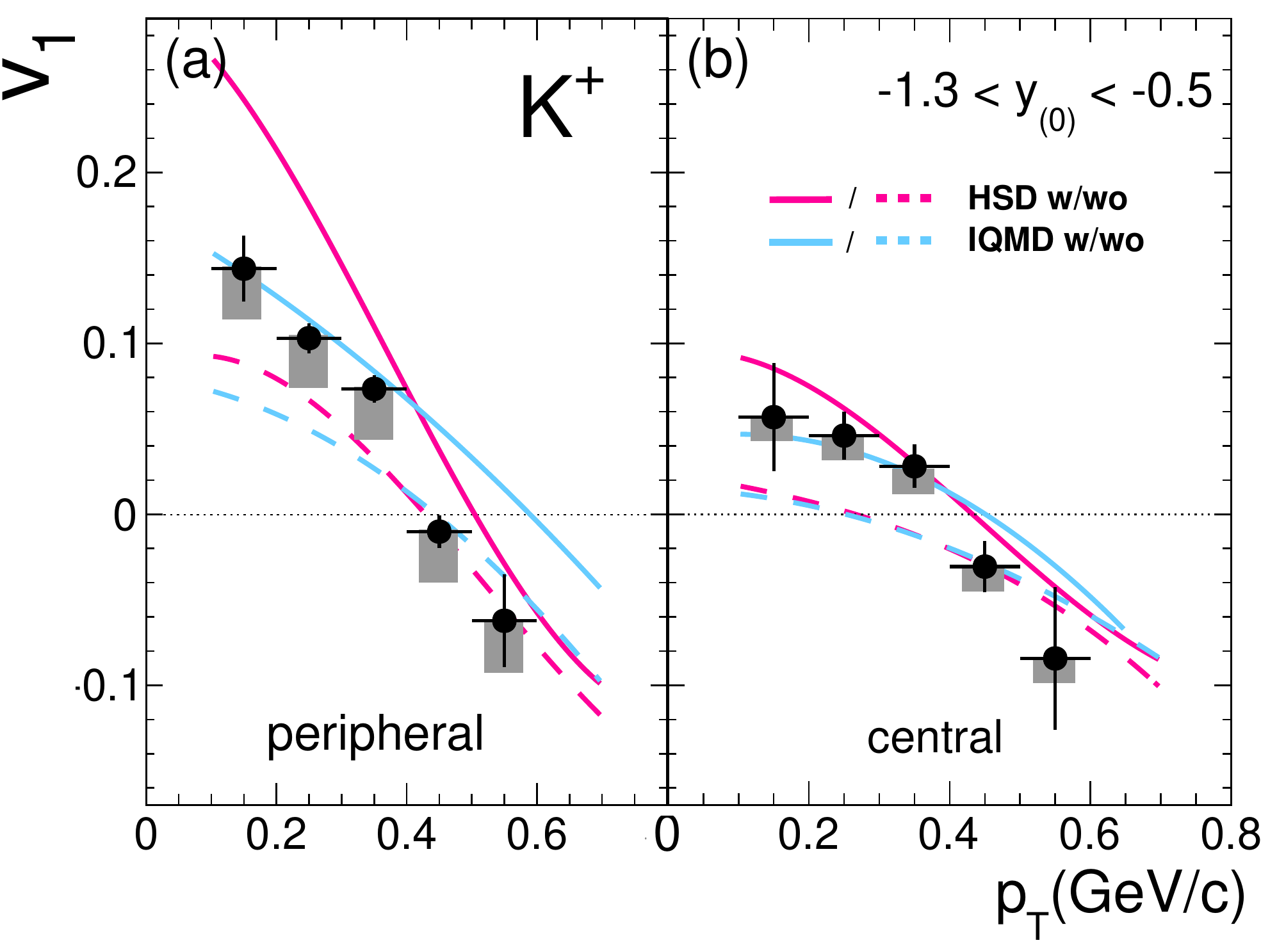}
 \caption{\label{fig:v1pt.cent} (color online) 
 Transverse momentum($p_{T}$) dependence of $v_{1}$ distributions 
 for  $ K^{+} $ mesons in peripheral (a) and central (b)
 collisions in comparison to  HSD and IQMD predictions with (solid lines) and without
 (dashed lines) in-medium potentials.
}
 \end{figure}

The experimental data on $ v_{1} $ and $ v_{2} $ of  $ K^{\pm} $  for  the
total event sample (a) (see Table~\ref{table:rpcor}) are presented as function of
 $y_{(0)}$ in Fig.~\ref{fig:v1} and Fig.~\ref{fig:v2}.  
$ v_{1}$ is by definition antisymmetric with respect to mid-rapidity,
therefore it should vanish at mid-rapidity for a symmetric collision system.
However, we observe, like in other FOPI data \cite{Crochet.2000}\cite{Reisdorf2012}, a deviation from zero. 
Systematic investigation of  $v_{1}$ at mid-rapidity $(y_{0}=0)$ showed a dependence on
centrality, particle type and system size suggesting a correlation with track density. 
When reconstructing the reaction plane with the best possible accuracy
(i.e. largest polar angle range $6.5^{\circ} < \theta_{lab} <113^{\circ}$)  
the  $v_{1}$ at mid-rapidity was found to be independent of transverse momentum in the system Ni+Ni at 2AGeV.
The origin of the $v_{1}$-`shift' was investigated by means of Monte Carlo study using the GEANT3 package \cite{geant}. 
Within the Monte Carlo framework the FOPI apparatus is described including the resolution in energy deposition and spatial position, 
front-end electronic processing, hit reconstruction, hit tracking and track matching between 
the sub-detectors. The output of GEANT was analyzed in the same way as the experimental data.
As observed in the data, the $v_{1}$-`shift' at mid-rapidity was found to depend on centrality and particle type, 
and to be independent of transverse momentum. However, the magnitude of the shift is underestimated by the simulation, 
and hence the Monte Carlo data can not be used to correct the experimental values quantitatively. 
We adopt a conservative approach and attribute the full shift at mid-rapidity as systematic uncertainty
(boxes in  Fig.~\ref{fig:v1} and Fig.~\ref {fig:v1pt.cent}). Since the local track density seen by the CDC only 
depends on the azimuthal angle and not on the polar angle we assume the systematic uncertainty to be rapidity independent.

The $ v_{1} $ values of $K^{-}$ 
are compatible with zero within the statistical sensitivity of the data (Fig.~\ref{fig:v1})
In order to reduce the statistical error, Fig.~\ref{fig:v1} also contains a 
$K^{-}$ data point with an upper momentum limit of $1.0$ GeV/$c$ (star symbol) from subset of runs with
 improved resolution.

Near target rapidity $K^{+}$ mesons show a collective in-plane deflection in
the direction opposed to that of protons (Fig.~\ref{fig:v1}). 
This pattern is called  `anti-flow' and is  in agreement with previous FOPI measurement \cite{Crochet.2000}. 
Additionally the  $K^{+}$ mesons  are observed  to collectively move out-of-plane (Fig.~\ref{fig:v2})
as indicated by the negative  $ v_{2} $ values. 
In case of $ K^{-} $ mesons (Fig.~\ref{fig:v1} (c) and Fig.~\ref{fig:v2} (c)), both
$v_{1} $ and $ v_{2} $ are compatible with zero
within the  statistical sensitivity of the data, i.e. an isotropic emission pattern is 
observed.

The KaoS collaboration has measured  $v_{2}$ coefficients
of $ K^{\pm} $ to be $v_{2}(K^+) = -0.05 \pm 0.03$ and  $v_{2}(K^-) = -0.09 \pm 0.06$ at mid-rapidity
for the same collision system at the same beam energy,  
however, with a different detector acceptance and collision  
centrality  ($3.8 < b_{geo} < 6.5$ fm) \cite{Uhlig.2005}.                                                            
The $v_{2}$ values 
from our data, reduced to the same centrality range and acceptance,  
are  compatible with the KaoS results, but do not show any indication for 
in-plane emission of $K^{-}$ mesons.

In order to link the flow measurements to the $ K^{\pm} $ properties in
the nuclear medium, a comparison to the predictions of transport model  calculations is necessary. 
For this analysis we utilize the Hadron String Dynamics (HSD)  model
\cite{Cassing.1999}  
and Isospin Quantum Molecular Dynamics (IQMD) \cite{hartnack.1998} 
offering a state-of-the-art description of kaon dynamics
\cite{Hartnack2011}.
The models employ different in-medium scenarios for the modification of strange particle
properties in the dense and hot medium: in HSD
 the chiral perturbation theory \cite{Mishra.2004} for kaons  and a coupled channel
G-matrix approach \cite{Cassing.2003} for antikaons are implemented. In IQMD 
transport approach the relativistic mean-field model for
kaons and antikaons based on a chiral SU(3) model is used \cite{SchaffnerBielich.1997}.

The centrality selection imposed on the data is realized by weighting the
events with an impact parameter dependent function. 
This function is obtained by evaluating the influence of a multiplicity selection 
on the impact parameter distribution within the IQMD model
which describes the multiplicity distribution -- after cluster formation -- reasonably well. 
Earlier data on flow of $K^{+}$ mesons \cite{Crochet.2000} and the $ K_{s}^{0} $ spectra
in pion induced reactions \cite{lotfi} were successfully described by HSD  with a 
repulsive $KN$ potential of $ 20\pm5 $~MeV  for particles at rest ($p=0$), 
at normal nuclear matter density and a linear dependence on baryon density. 
Employing this parametrization for the  $K^{+}N$ potential in both HSD and IQMD, and a similar,
but attractive one with $U_{K^{-}N}(\rho=\rho_{0}, p=0) = -45$\,MeV  in IQMD 
and a G-Matrix formalism corresponding to $U_{K^{-}N}(\rho=\rho_{0}, p=0) = -50$\,MeV in HSD for the
$K^{-}N$ potential 
 the model predictions depicted by
the full lines in Fig.~\ref{fig:v1} and Fig.~\ref{fig:v2} are obtained.
The phase space distributions obtained from the transport calculations are filtered 
for the detector acceptance. 
The flow observables are calculated using the true reaction plane.
Typical statistical uncertainties in the calculations are of the order 
$\Delta v_1 \approx 0.005$ and $\Delta v_2 \approx 0.01$. 
The effect of the in-medium potentials is visible in the difference to the 
model calculations without in-medium potential (dashed lines) that 
still include $K^{\pm}N$ rescattering and absorption processes for $ K^{-}$ mesons.

Inspection of Fig.~\ref{fig:v1} reveals that according to the transport models the
largest sensitivity to the presence of in-medium potentials is achieved with the
sideflow observable,  $v_{1}$,  near target rapidity.
Without any in-medium modifications the $K^{+}$ mesons should be emitted nearly 
isotropic, i.e. $v_{1}$ and $ v_{2} $ values are close to zero. The presence of a 
repulsive  $K^{+}N$ potential manifests itself by pushing the
$K^{+}$ mesons away from the protons, thus generating the `anti-flow' signature 
of $K^{+}$ mesons.
The magnitude of the `anti-flow' is correctly described by IQMD transport approach with the
 assumption of a $K^{+}N$ potential of $ 20\pm5 $~MeV, while HSD predicts the `anti-flow' 
effect but quantitatively overestimates the magnitude of the experimentally observed `anti-flow'.
For $K^{-}$ mesons the interpretation is different: because of the
strong absorption due to strangeness exchange reactions with baryons, an
`anti-flow' signature is expected without the presence of a potential (Fig.~\ref{fig:v1}). 
This is clearly disfavored by the data. 
Assuming an additional attraction of $K^{-}$ toward protons, due to strong interaction, the
 IQMD transport model predicts an almost isotropic emission pattern, as it is observed in the data.
 HSD predicts a strong `flow' signature in the near target region, though.
Following both model predictions the data indicate the presence 
of an attractive $K^{-}N$ potential. Following the IQMD approach,
the depth of the potential can be constrained to
 $U_{K^{-}N}(\rho=\rho_{0}, p=0) = -40 \pm 10$\,MeV.

The second harmonic $v_{2}$  of $K^{+}$ mesons  (Fig.~\ref{fig:v2})
shows a squeeze-out signature at mid-rapidity that is explained within the IQMD
model by the presence of an in-medium potential. The HSD model predicts 
a weak squeeze-out effect, deviating by about 2 $\Delta v_2$ from the data.
Within the statistical sensitivity of the data this observable 
does not show any sensitivity to the potential away from mid-rapidity.

In the near target rapidity region the experimental $v_{2}$ is underestimated by both model 
calculations.
However, the deviation is at the limit of the statistical significance. Note that also the
$v_{2}$ values of protons  (Fig.~\ref{fig:v2}) are not reproduced by HSD.
In the case of $K^{-}$ elliptic flow, experimental
uncertainties are too large to draw any conclusion about the  $K^{-}N$ potential.  

To probe the consistency of the transport model description of the
current data, shown in Fig.~\ref{fig:v1},  we present in
Fig.~\ref{fig:v1pt.cent} the  differential dependence of $v_{1}$ on the 
transverse momentum  $p_{T}$  for
$K^{+}$ mesons near target rapidity ($-1.3<y_{(0)}<-0.5$)
for the two centrality classes (p) and (c) defined in Table~\ref{table:rpcor}.

In the central event sample the data are compatible with both, HSD and IQMD,
calculations employing the in-medium potential described above, in agreement to
previously published  FOPI results  \cite{Crochet.2000}.
The IQMD calculations reproduce the transverse momentum dependence 
and the strength of the $v_{1}$ coefficient for low transverse momenta ($p_{T}<0.4$ GeV/c).
This quality of IQMD is also observed for the peripheral event sample, where
the data show a slightly stronger $p_{T}$ dependence as compared to the
central case. Within HSD the transverse momentum dependence is
strongly over-predicted leading to very large asymmetries at small $p_{T}$ that
are excluded by the data. 

The influence of the Coulomb interaction was studied within the HSD transport model by comparing 
the flow patterns of $K^0_S$  and $K^+$ mesons. Both members of the isospin
 doublet show a similar $p_{T}$ dependence of $v_{1}$, but in case of $K^+$ mesons
the predicted `anti-flow' is up to 12 \% higher at low transverse momenta.
This difference is attributed to the additional repulsion due to electromagnetic interaction 
between $K^+$ mesons and protons in the near target region. The long range influence 
of the Coulomb attraction/repulsion between kaons and nucleons of the projectile and target remnants 
was investigated with the SACA clusterization algorithm  \cite{saca} which simulates 
the propagation of particles in the Coulomb field up to flight times of $\sim$ 10,000 fm/c. 
 Statistically significant influence was
 found only for the very small momenta, beneath the detector acceptance.
 We conclude that most of the asymmetry is caused by the strong interaction and that $v_{1}$ 
can constrain the depth of the $KN$ potential.
\section{Conclusions}
We have measured the azimuthal emission patterns of
$K^{\pm}$ mesons  in heavy-ion collisions near the strangeness production threshold energies.
In case of $K^{+}$  mesons a weak  in-plane `anti-flow'  with respect to protons   and
a slight `squeeze-out'  are observed. 
For $K^{-}$  mesons, within large statistic uncertainties,
isotropic emission pattern is observed. Despite the large uncertainties, 
the comparison to two independent predictions of HSD and IQMD  without potential especially
of the first Fourier coefficient implies the
existence of a weakly attractive $K^{-}N$ in-medium potential. 
Furthermore, the IQMD transport approach, which is able to
reproduce the dynamics of nucleons and $K^{+}$ mesons, 
suggests a $K^{-}N$ in-medium potential of $U_{K^{-}N} = -40 \pm 10$\,MeV.

The theoretical modeling of the in-medium potentials, or more generally of the
in-medium interactions, is reasonably well achieved within the IQMD transport approach
as is demonstrated by the detailed comparison of the differential flow pattern
of the $K^{+}$ mesons. Within HSD, the general dynamics of nucleons and $K^{+}$ mesons is reproduced as well,
 however the quantitative description of the flow pattern in some regions of the phase space
is not accurately achieved yet. 
The observed dependencies and
sensitivities point to the feasibility to extract the strength of the in-medium potentials
from a quantitative description of a complete set of flow data. 
More systematic data and theoretical efforts are clearly necessary to reach this important goal.
\begin{acknowledgments}
This work was supported by  BMBF-05P12VHFC7,  
the KOSEF(F01-2006-000-10035-0), 
by BMBF-05P12RFFCQ, 
by the Polish Ministry of Science and Higher Education(DFG/34/2007), 
the agreement between GSI and IN2P3/CEA, the HIC for FAIR, 
the Hungarian OTKA (71989), 
by NSFC (project 11079025),
by DAAD (PPP D/03/44611), 
by DFG (Projekt 446-KOR-113/76/04)
and by the EU, 7th Framework Program, Integrated Infrastructure: Strongly Interacting
Matter (Hadron Physics), Contract No. RII3-CT-2004-506078.
\end{acknowledgments}

\end{document}